\documentclass[12pt]{iopart}

\usepackage{graphicx}
\usepackage{color}
\def\prl{Phys. Rev. Lett.}
\def\prd{Phys. Rev. D}
\def\cqg{Class. Quantum Grav.}

\def\apj{Astrophys. J.}
\def\apjs{Astrophys. J. Suppl.}

\def\mnras{Mon. Not. R. Astron. Soc.}

\begin{document}

\title{Bondi Accretion in Trumpet Geometries}

\author{August J. Miller and Thomas W. Baumgarte}
\address{Department of Physics and Astronomy, Bowdoin College, Brunswick, ME 04011, USA}
\ead{augustm2@illinois.edu, tbaumgar@bowdoin.edu}

\begin{abstract}
The Bondi solution, which describes the radial inflow of a gas onto a non-rotating black hole, provides a powerful test for numerical relativistic codes.  However, the Bondi solution is usually derived in Schwarzschild coordinates, which are not well suited for dynamical spacetime evolutions.  Instead, many current numerical relativistic codes adopt moving-puncture coordinates, which render black holes in trumpet geometries.  Here we transform the Bondi solution into trumpet coordinates, which result in regular expressions for the fluid flow extending into the black-hole interior.  We also evolve these solutions numerically and demonstrate their usefulness for testing and calibrating numerical codes.
\end{abstract}


\section{Introduction}
\label{sec:intro}

Accretion processes are ubiquitous in astrophysics.  Of particular importance in relativistic astrophysics is accretion onto black holes.  In general, such processes are quite complicated and require a numerical treatment (see, e.g., \cite{AbrF13} for a recent review), but a simple analytical solution, describing the spherically symmetric, radial flow of a low-density fluid onto a non-rotating black hole, was derived in 1952 by Hermann Bondi \cite{Bon52}.  Both his original solution, which was derived in Newtonian gravity, and its relativistic analogues (see, e.g., \cite{Mic72,Beg78}, as well as Appendix G in \cite{ShaT83} for a textbook treatment), are commonly referred to as Bondi solutions.   Bondi solutions have also been generalized to include radiative processes \cite{Sha73a} as well as magnetic fields \cite{Sha73b,DeVilH03}.  The effects of self-gravity, for fluids whose density is sufficiently large so that its effects on the spacetime metric can no longer be neglected, have been studied by \cite{Mal99,KarKMMS06,MacM08}.   These latter solutions have also been generalized  to include radiative \cite{MalR10} and cosmological effects \cite{KarM13,MacM13,MacMK13,BabDE13}.  Homoclinic orbits in the accretion flow, in addition to the standard transonic solutions, have been discussed in \cite{Mac15,ChaMS15}.

Even though the Bondi solutions describe a very special scenario, they play an important role in our understanding of accretion processes.  On the one hand, they illustrate some general features of accretion flow in the context of a simple solution; on the other hand, they provide a powerful test for relativistic hydrodynamics or magnetohydrodynamics codes designed, for example, to simulate more general accretion processes.  It is the latter aspect that we will focus on in this paper.

The Bondi solution has been used as a test case for many numerical codes (see \cite{HawSW84a,HawSW84b,DeVilH03,GamMT03,DueLSS05b,FabBEST07,MoeMFHNBLORS14,MelMOPRZ16} for some examples).  One complication arises from the fact that the Bondi solution is usually given in Schwarzschild coordinates, which, in general, are not well suited for numerical simulations.  In numerical codes that assume a fixed spacetime geometry, these problems can be solved by using, for example, Kerr-Schild (ingoing Eddington-Finkelstein coordinates), which avoid the problems associated with the black-hole event horizon in Schwarzschild coordinates.    However, in codes that do evolve the spacetime dynamically and self-consistently with matter sources, even these coordinates are inadequate for dealing with the problems associated with the spacetime singularity (e.g.~\cite{BauS10}).  In spherical symmetry, Lagrangian (or co-moving) coordinates have also proven useful, but they are difficult to employ in multidimensional codes.

Self-consistent evolutions of black-hole spacetimes, without the assumption of symmetries, became possible with the simulations of \cite{Pre05b} and \cite{CamLMZ06,BakCCKM06a}.  The latter two groups adopted a now quite commonly used approach, in which Einstein's equations, expressed in the BSSN formulation \cite{NakOK87,ShiN95,BauS98} or some variation, are evolved in so-called {\em moving-puncture} coordinates.  These coordinates, which consist of a ``1+log'' slicing condition for the lapse function \cite{BonMSS95} and a ``Gamma-driver'' condition for the shift vector \cite{AlcBDKPST03}, bring the spatial slices of black holes into a so-called trumpet geometry (see \cite{HanHPBO06,HanHOBO08}, as well as \cite{BauS10} for a textbook treatment; a simple analytical example can be found in \cite{DenB14}).  Summarized briefly, trumpet slices penetrate the event horizon of the black hole smoothly and terminate on a limiting surface of areal radius greater than zero, thereby shielding the simulation from the effects of the curvature singularity.

One method of using the Bondi solution as a test case for a code that adopts moving-puncture coordinates is to express the initial data in isotropic coordinates on a slice of constant Schwarzschild time (see, e.g., \cite{FabBEST07}).  During the evolution, the solution will undergo a coordinate transition and eventually settle down into a trumpet geometry.  While workable, this approach has several disadvantages.  In particular, because of the coordinate transition, only gauge-invariant quantities can be compared directly with the analytical Bondi solution.  Moreover, isotropic coordinates on slices of constant Schwarzschild time cover only the exterior of the black hole, so the region near and inside the black hole has to be initialized with artificial data (see \cite{EtiFLSB07,BroSSTDHP07} for examples).

In this paper we demonstrate that it is easy to transform the Bondi solution from Schwarzschild coordinates into trumpet coordinates, i.e., isotropic coordinates in a trumpet geometry.  Casting the Bondi solution in trumpet coordinates avoids the coordinate transition when the data are evolved using the moving-puncture method, resulting in a time-independent solution.  This implies that all evolved quantities, including those that are gauge-dependent, can be compared directly with the analytical solution.  Moreover, the Bondi solution in these coordinates extends smoothly into the black-hole interior, eliminating the need to initialize the interior with artificial data.

Our paper is organized as follows.  In Section \ref{sec:review} we present a brief review of the Bondi solution in Schwarzschild coordinates.  Then, in Section \ref{sec:trans}, we describe how this solution can be transformed into other coordinate systems and present several numerical examples, demonstrating the usefulness of the transformed Bondi solution as a code test.  We conclude with a brief summary in Section \ref{sec:summary}.

\section{Bondi accretion in Schwarzschild coordinates}
\label{sec:review}

The relativistic Bondi solution is usually expressed in Schwarzschild coordinates, in which the line element takes the form
\begin{equation}
ds^2 = -\left(1 - \frac{2M}{R}\right)dT^2 + \left(1 - \frac{2M}{R}\right)^{-1}dR^2 + R^2 d\Omega^2.
\end{equation}
Here $M$ is the mass of the black hole, $T$ is the Schwarzschild time, and $R$ is the Schwarzschild, or areal, radius.  A derivation of the Bondi solution in these coordinates can be found, for example, in Appendix G of \cite{ShaT83}; here we list only the most important equations.

We consider spherically symmetric, radial fluid flow onto a non-rotating black hole of mass $M$.  We will assume the density of the fluid to be sufficiently small such that we can neglect its self-gravity and approximate $M$ to remain constant, in which case the solution can be given in analytical form.  Since the flow is purely radial, we may write the fluid four-velocity as $u^a = \big(u^t, u^R, 0, 0\big)$.  For convenience, we will refer to the negative radial component of the four-velocity as $u$ in all coordinate systems; in the Schwarzschild coordinates of this Section we have $u = - u^R$.
Since we are considering accretion solutions only, we will have $u \geq 0$ always.  We further assume that the fluid is at rest, $u = 0$, at spatial infinity.

We can derive the relativistic Bondi solution from two fundamental equations, namely the law of baryon conservation,
\begin{equation} \label{baryon_cons}
\nabla_a\left(\rho_0 u^a\right) = 0,
\end{equation}
where $\rho_0$ is the proper rest-mass density, and the conservation of energy-momentum,
\begin{equation} \label{se_cons}
\nabla_b T^{ab} = 0.
\end{equation}
Here
\begin{equation}
T^{ab} = \left(\rho + P\right)u^a u^b + Pg^{ab}
\end{equation}
is the stress-energy tensor for a perfect fluid, $\rho$ the total energy density, and $P$ the pressure.  Evaluating the spatial part of Eq.~(\ref{se_cons}) yields the relativistic Euler equation,
\begin{equation} \label{euler}
\left(\rho + P\right)u^b\nabla_b u^a = -\partial^a P - u^a u^b\partial_b P,
\end{equation}
while the time component yields the first law of thermodynamics under the condition of constant entropy,
\begin{equation} \label{entropy}
d\left(\frac{\varepsilon}{\rho_0}\right) = -Pd\left(\frac{1}{\rho_0}\right),
\end{equation}
where $\varepsilon = \rho - \rho_0$ is the internal energy density.  We therefore conclude that the flow must be adiabatic in the absence of shocks.  In addition to Eqs.~(\ref{baryon_cons}) and (\ref{se_cons}), we adopt a Gamma-law equation of state,
\begin{equation} \label{eos}
P = \left(\Gamma - 1\right)\varepsilon,
\end{equation}
where $\Gamma$ is the adiabatic index.  For adiabatic flow, this implies the polytropic relation
\begin{equation} \label{polytropic}
P = \kappa\rho_0^\Gamma,
\end{equation}
where $\kappa$ is the polytropic constant.  Finally, the speed of sound is given by
\begin{equation}
a = \left(\frac{dP}{d\rho}\right)^{1/2} = \left(\frac{\Gamma P}{\rho + P}\right)^{1/2}.
\end{equation}

In Schwarzschild coordinates, the law of baryon conservation (\ref{baryon_cons}) and the relativistic Euler equation (\ref{euler}) can be integrated to obtain the accretion rate equation,
\begin{equation} \label{acc_rate}
4\pi\rho_0 uR^2 = \mathrm{constant} = \dot{M},
\end{equation}
and the relativistic Bernoulli equation,
\begin{equation} \label{bernoulli}
\left(\frac{\rho + P}{\rho_0}\right)^2\left(1 - \frac{2M}{R} + u^2\right) = \mathrm{constant} = \left(\frac{\rho_\infty + P_\infty}{\rho_{0,\infty}}\right)^2,
\end{equation}
respectively.  Here and in the following we denote quantities at spatial infinity with a subscript $\infty$.  In addition, all smooth solutions to the conservation equations (\ref{baryon_cons}) and (\ref{euler}) must pass through a critical point at which
\begin{equation}
u_s^2 = \frac{M}{2R_s}
\end{equation}
and
\begin{equation}
a_s^2 = \frac{u_s^2}{1 - 3u_s^2} = \frac{M}{2R_s - 3M}.
\end{equation}
Here $R_s$ is the critical radius, i.e., the radius at which the flow passes through the critical point, and $u_s$ and $a_s$ are the fluid velocity and sound speed at $R = R_s$.  In the Newtonian limit, when $R_s \gg M$, the critical radius coincides with the sonic radius, at which $u_s = a_s$.

\begin{table}
\begin{center}
\begin{tabular}{ccc | ccccc}
\hline\hline
$\Gamma$ & $R_s$ & $\dot{M}$ & $u_s$ & $a_s$ & $\rho_{0,s}$ & $\kappa$ & $E$ \\
\hline
$4/3$ & $10M$ & $10^{-4}$ & 0.2236 & 0.2425 & $3.559 \times 10^{-7}$ & 7.560 & 1.253 \\
\hline\hline
\end{tabular}
\end{center}
\caption{Values of key fluid parameters for the Bondi solution used in our numerical simulations, which is characterized by $\Gamma = 4/3$, $R_s = 10M$, and $\dot{M} = 10^{-4}$.  (We take $M = 1$ for simplicity.) Here $\rho_{0,s}$ is the rest-mass density at the critical radius and $E = \left[\left(\rho_\infty + P_\infty\right)/\rho_{0,\infty}\right]^2$ is the Bernoulli constant.}
\label{Tab1}
\end{table}

We can specify a unique solution to the equations of Bondi accretion by choosing the values of four parameters.  In all of our numerical simulations, we use the Bondi solution characterized by $\Gamma = 4/3$, $\dot{M} = 10^{-4}$, $R_s = 10M$, and $M = 1$.  We list the values of key fluid parameters for this solution in Tab.~\ref{Tab1}.  Once we have chosen our parameter values, we combine the accretion rate equation (\ref{acc_rate}) and the Bernoulli equation (\ref{bernoulli}) to obtain a nonlinear equation for the rest-mass density $\rho_0$ only, using the equation of state (\ref{eos}) and polytropic relation (\ref{polytropic}) to eliminate dependence on all other fluid variables.  We solve this equation iteratively for $\rho_0$ at discrete radial values $R$ and then solve for $u$ at each $R$ using Eq.~(\ref{acc_rate}).  We thereby obtain solutions $\rho_0\left(R\right)$ and $u\left(R\right)$ for all desired values for $R$.

We also compute the time-component of the four-velocity $u^t$ and the normal three-velocity $v \equiv -v^R$ as functions of $R$; see Eqs.~(\ref{u_t}) and (\ref{v_r}) below.  In Schwarzschild coordinates, these quantities become singular on the black-hole horizon.

\begin{figure}
\centering
\includegraphics[width=0.75\textwidth]{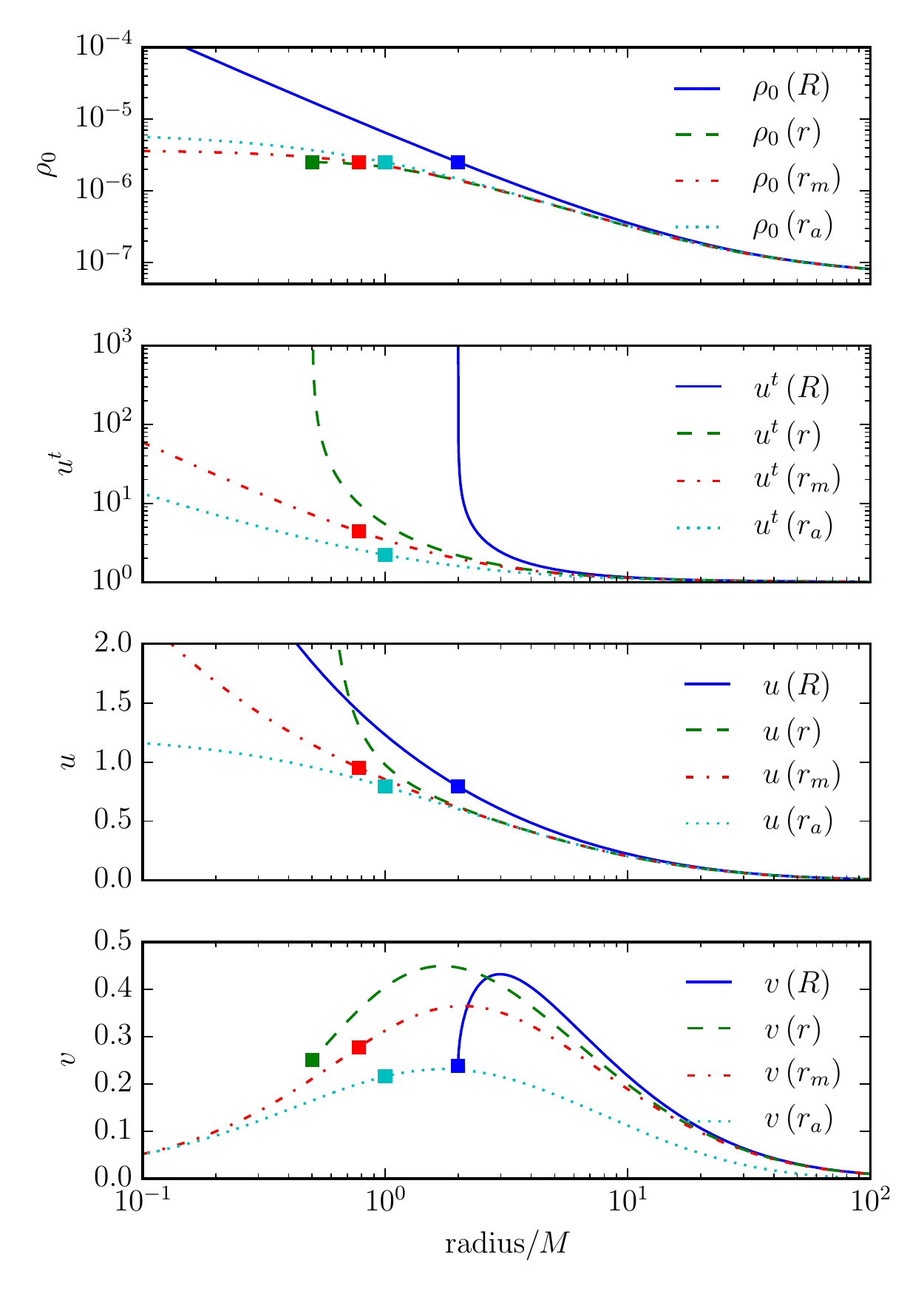}
\caption{Profiles of the fluid variables $\rho_0$, $u^t$, $u = -u^r$, and $v = -v^r$ for the Bondi solution characterized by $\Gamma = 4/3$, $\dot{M} = 10^{-4}$, and $R_s = 10M$ in the four different coordinate systems considered in this paper, namely Schwarzschild coordinates (radius $R$, solid lines), isotropic coordinates on a slice of constant Schwarzschild time $T$ (radius $r$, dashed lines), isotropic coordinates in a maximal trumpet geometry (radius $r_m$, dot-dashed lines), and isotropic coordinates in an analytical trumpet geometry (radius $r_a$, dotted lines).  Boxes mark the location of the event horizon in each coordinate system.  For the trumpet geometries all quantities extend smoothly into the black-hole interior.}
\label{Fig1}
\end{figure}

In Fig.~\ref{Fig1} we show radial profiles of $\rho_0$, $u^t$, $u$, and $v$ in Schwarzschild coordinates (solid lines), along with those in all other coordinate systems considered in the following sections, namely isotropic coordinates on a slice of constant Schwarzschild time, in a maximal trumpet geometry, and in an analytical trumpet geometry.  Fig.~\ref{Fig1} highlights one of the main results of this paper, namely that in trumpet geometries all fluid variables extend smoothly into the black-hole interior.

\section{Transformation to other coordinate systems}
\label{sec:trans}

\subsection{General expressions}
\label{sec:trans:general}

We consider transformations from Schwarzschild coordinates to other spherically symmetric coordinate systems that are both time-independent and spatially isotropic.  The metric can thus be written as
\begin{equation} \label{metric}
ds^2 = \left(-\alpha^2 + \psi^4 \beta^2\right)dt^2 + 2\beta dtdr + \psi^4\left(dr^2 + r^2 d\Omega^2\right),
\end{equation}
where $\alpha$ is the lapse function, $\beta^i = (\beta,0,0)$ is the shift vector, and $\psi$ is a conformal factor.  The normal vector $n^a$ on slices of constant coordinate time $t$ is given by
\begin{equation}
n^a = \frac{1}{\alpha}\left(1,-\beta,0,0\right),
\end{equation}
and the extrinsic curvature $K_{ij}$ can be computed from
\begin{equation}
K_{ij} = \frac{1}{2 \alpha}\left(D_i\beta_j + D_j\beta_i\right),
\end{equation}
where $D_i$ is the covariant derivative associated with the spatial metric $\gamma_{ij} = \psi^4\eta_{ij}$.

Transformation of the Bondi solution to such a coordinate system can be performed in two steps.  We first transform to a new time coordinate $t$, which allows for a different slicing of the Schwarzschild spacetime, while keeping the Schwarzschild radial coordinate $R$ as our radial coordinate.  This transformation can be accomplished with the help of a height function
\begin{equation} \label{heightfunction}
t = T + h\left(R\right).
\end{equation}
Since this transformation does not affect the spatial coordinates, and since all quantities are independent of time, all covariant spatial components of tensors, as well as all scalars, remain unchanged.  (It can also be shown formally that the form of the fluid equations is invariant under this transformation.)  

In the second step, we transform to a new (isotropic) radial coordinate $r$ within the new spatial slice.  Under this transformation, the radial component of the four-velocity transforms as
\begin{equation} \label{u_r}
u^r = \frac{\partial r}{\partial R}u^R.
\end{equation}
We also compute the three-velocity of the fluid as measured by a normal observer,
\begin{equation} \label{v_r}
v^r \equiv \frac{1}{W}{\gamma^r}_a u^a = \frac{1}{\alpha}\left(\frac{u^r}{u^t} + \beta\right),
\end{equation}
which is used in many formulations of relativistic hydrodynamics.  Here $W \equiv -n_a u^a = \alpha u^t$ is the Lorentz factor between a normal observer and an observer comoving with the fluid.  In order to compute $v^r$, we need the time component of the fluid four-velocity $u^t$ in addition to the radial component $u^r$.  One method of computing $u^t$ would be to start with the result in Schwarzschild coordinates and perform the appropriate coordinate transformations.  However, since the flow is radial, we can also calculate $u^t$ from the normalization of the four-velocity, $u_a u^a = -1$.  Using the metric (\ref{metric}), we find
\begin{equation} \label{u_t}
u^t = \frac{1}{\alpha^2 - \psi^4\beta^2}\left[-\psi^4\beta u + \sqrt{\psi^8\beta^2 u^2 + \left(\alpha^2 - \psi^4\beta^2\right)\left(\psi^4 u^2 + 1\right)}\right],
\end{equation}
where $u = - u^r$.  As a scalar, the rest-mass density $\rho_0$ is invariant under both the time and spatial transformations.



We can now compute the Bondi solution in a given coordinate system as follows.  We first compute the fluid variables $\rho_0\left(R\right)$ and $u^R\left(R\right)$ in Schwarzschild coordinates, as described in Section \ref{sec:review}.  We then find $u^r$ from Eq.~(\ref{u_r}) and $u^t$ from Eq.~(\ref{u_t}), and express the radial dependence in terms of $r$ rather than $R$.  Finally, we insert these quantities into Eq.~(\ref{v_r}) to find $v^r$, which completes the prescription of the Bondi solution.

\subsection{Isotropic coordinates on slices of constant Schwarzschild time}
\label{sec:trans:iso}

Before transforming to trumpet coordinate systems in the following sections, we first evaluate the expressions of Section \ref{sec:trans:general} for isotropic coordinates on slices of constant Schwarzschild time $T$ (i.e., for which the height function in (\ref{heightfunction}) vanishes, $h = 0$, so $t = T$).  In terms of the areal radius $R$, the isotropic radius $r$ in these coordinates is given by
\begin{equation}  \label{r_of_R_iso}
r = \frac{1}{2}\left[R - M \pm \sqrt{R\left(R - 2M\right)}\right],
\end{equation}
which places the event horizon at $r = M/2$.  The lapse is given by
\begin{equation} \label{lapse_iso}
\alpha = \frac{1 - M/\left(2r\right)}{1 + M/\left(2r\right)},
\end{equation}
the conformal factor by
\begin{equation} \label{psi_iso}
\psi = 1 + \frac{M}{2r},
\end{equation}
and the shift and extrinsic curvature both vanish ($\beta = 0 = K_{ij}$).  We note that these coordinates cover only the black-hole exterior, $R > 2M$.

From Eqs.~(\ref{u_r}) and (\ref{u_t}) we find
\begin{equation} \label{u_r_iso}
u^r = \frac{u^R}{\left(1 + M/\left(2r\right)\right)\left(1 - M/\left(2r\right)\right)}
\end{equation}
(compare Eq.~(B14) in \cite{FabBEST07}) and
\begin{equation} \label{u_t_iso}
u^t = \frac{1 + M/\left(2r\right)}{1 - M/\left(2r\right)}\left[\left(1 + \frac{M}{2r}\right)^4 u^2 + 1\right]^{1/2},
\end{equation}
respectively.  Inserting the above expressions into (\ref{v_r}) yields
\begin{equation} \label{v_r_iso}
v^r = -u\left[\left(1 + \frac{M}{2r}\right)^4 u^2 + 1\right]^{1/2}.
\end{equation}
Results for the fluid variables in isotropic Schwarzschild coordinates are included in Fig.~\ref{Fig1}.  We note that the expressions for $u^t$ and $u^r$ become singular at the horizon, which is marked in the figure by a solid box.

In order to construct Bondi initial data in these coordinates, the authors of \cite{FabBEST07} replaced the exact data with a fitting function inside a radius $r = M$.  While these artificial data do not satisfy Einstein's constraint equations, the numerical effects of this discrepancy are quite small, since this part of the fluid is absorbed into the black hole very quickly.  Here we use the same prescription to regularize the initial data close to the black hole.  In \cite{FabBEST07} the initial data were then evolved using moving-puncture coordinates, which results in a coordinate transition from isotropic coordinates on a slice of constant Schwarzschild time to isotropic coordinates in a trumpet geometry.  As a consequence, only gauge-invariant quantities, for example the rest-mass density $\rho_0$ plotted as a function of $R$, can be compared directly with the analytical solution; all gauge-dependent quantities will exhibit some form of time-dependence.

In order to demonstrate this effect, we perform numerical simulations using as initial data the expressions for the Bondi solution in isotropic coordinates on a slice of constant Schwarzschild time $T$.  We evolve these data with a code that implements the BSSN formulation of Einstein's equations \cite{NakOK87,ShiN95,BauS98}, together with the equations of relativistic hydrodynamics, in spherical polar coordinates (see \cite{BauMCM13,MonBM14,BauMM15}).  One ingredient in these implementations of the gravitational field equations is the usage of a so-called reference metric \cite{Bro09,Gou12}.  As discussed in \cite{MonBM14,BauMM15}, this technique can also be used in the equations of relativistic hydrodynamics.  We distinguish between a ``full'' approach, in which all fluid equations are expressed in terms of a reference metric, and a ``partial'' approach, in which the reference metric is used only in the relativistic Euler equation.  Both approaches have their respective advantages and disadvantages (see, e.g., \cite{BauMM15}); here we focus on the partial approach, which, in the case of Bondi flow, appears to lead to smaller errors in the fluid variables in the immediate vicinity of the black-hole puncture.  We solve the equations of relativistic hydrodynamics using a high-resolution shock-capturing scheme, employing a second-order slope limiter reconstruction scheme, namely the monotonic centered limiter \cite{Van77}, as well as the Harten-Lax-van Leer-Einfeld approximate Riemann solver \cite{HarLL83,Ein88}.  Our code does not make any symmetry assumptions, but we run it here with the minimum number of grid points in the angular directions, as is appropriate for a spherically symmetric spacetime.

\begin{figure}
\centering
\includegraphics[width=0.75\textwidth]{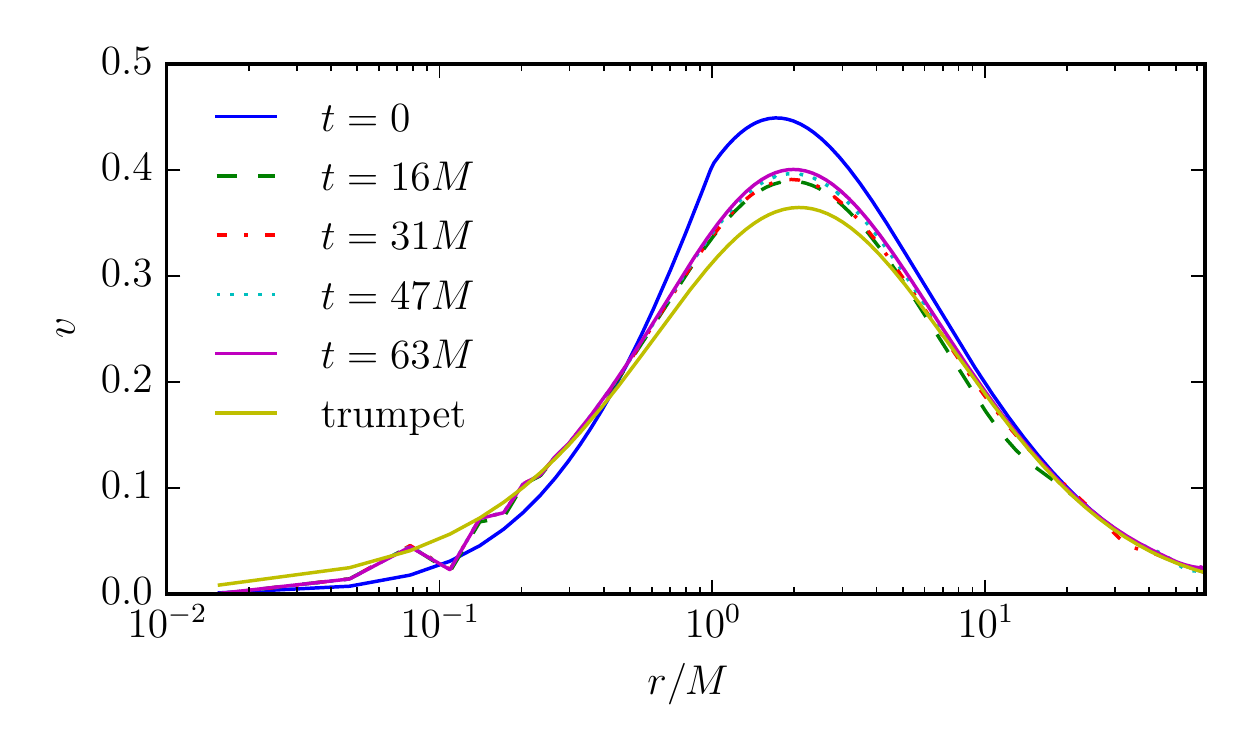}
\caption{The radial three-velocity $v = -v^r$ for isotropic initial data on a slice of constant Schwarzschild time $T$, evolved with the 1+log slicing condition (\ref{1+log}).  We used $N_r = 2048$ uniform radial grid points in this simulation, and imposed the outer boundary at $r_{\mathrm{max}} = 64M$.   The lines connect individual grid points; the ``choppy" behavior at small radii at $t > 0$ is caused by numerical error due to finite-differencing across the black hole puncture, where the conformal factor $\psi$ diverges (see Section \ref{sec:trans:max} for more details).  Inside $r = M$, the initial data at $t = 0$ are given by the artificial fitting functions suggested by \cite{FabBEST07}, which result in the kink visible in the figure.  Profiles at different instants of time can be clearly distinguished, demonstrating the time-dependence of the solution. In addition, the dynamical evolution does not settle down to the analytical solution for maximal trumpet data, since the spatial metric does not remain isotropic during the evolution.}
\label{Fig2}
\end{figure}

We adopt moving-puncture coordinates consisting of a (non-advective) 1+log slicing condition \cite{BonMSS95}
\begin{equation} \label{1+log}
\partial_t\alpha = -2\alpha K,
\end{equation}
together with a Gamma-driver shift condition \cite{AlcBDKPST03}.  In Eq.~(\ref{1+log}), $K \equiv \gamma^{ij}K_{ij}$ is the trace of the extrinsic curvature.  We start with a ``pre-collapsed'' lapse $\alpha = \psi^{-2}$ and zero shift ($\beta = 0$).  As expected, these coordinate conditions lead to a coordinate transition from the initial coordinates (isotropic coordinates on a slice of constant Schwarzschild time) to a trumpet geometry.  In Fig.~\ref{Fig2} we show snapshots of the fluid three-velocity $v = -v^r$ at different instants of coordinate time, evolved with $N_r = 2048$ uniform radial grid points and the outer boundary imposed at $r_{\mathrm{max}} = 64M$.  The fact that these profiles change in time is a consequence of the coordinate transition.

Evidently, isotropic coordinates on a slice of constant Schwarzschild time are inconvenient for numerical tests for two primary reasons.  First, these coordinates become singular on the event horizon and do not extend into the black-hole interior, necessitating the use of artificial initial data in the vicinity of the black hole; and second, they result in a coordinate transition when evolved with the moving-puncture method.  As we will show in the following two sections, casting the Bondi solution in trumpet coordinates avoids both of these problems.


\subsection{Maximally sliced trumpet coordinates}
\label{sec:trans:max}

As pointed out by \cite{HanHOBGS06}, the slicing condition (\ref{1+log}) will lead to maximal slicing ($K = 0$) once equilibrium with $\partial_t\alpha = 0$ has been reached.  Moreover, the corresponding maximal slice of Schwarzschild can also be transformed into isotropic coordinates, albeit only in parametric form \cite{BauN07}.  Specifically, the isotropic radius $r$ is then given by
\begin{equation} \label{r_of_R_max}
\begin{array}{rcl}
r & = & \displaystyle \left[\frac{2R + M + \sqrt{4R^2 + 4MR + 3M^2}}{4}\right] \\
& & \displaystyle \times \left[\frac{\big(4 + 3\sqrt{2}\big)\left(2R - 3M\right)}{8R + 6M + 3\sqrt{8R^2 + 8MR + 6M^2}}\right]^{1/\sqrt{2}}.
\end{array}
\end{equation}
In terms of $r$, the event horizon is now located at approximately $r = 0.779M$; evidently, the isotropic radius on this maximal slice is different from that on the slices of constant Schwarzschild time $T$ discussed in Section \ref{sec:trans:iso}.  Spatial slices terminate at $r = 0$, which corresponds to an areal radius $R = 3M/2$.  The lapse function is given by
\begin{equation} \label{lapse_max}
\alpha = \left(1 - \frac{2M}{R} + \frac{27M^4}{4R^4}\right)^{1/2},
\end{equation}
the shift by
\begin{equation} \label{shift_max}
\beta = \frac{3\sqrt{3}M^2}{4}\frac{r}{R^3},
\end{equation}
where $r$ is given above, and the conformal factor by
\begin{equation} \label{psi_max}
\begin{array}{rcl}
\psi & = & \displaystyle \left[\frac{4R}{2R + M + \sqrt{4R^2 + 4MR + 3M^2}}\right]^{1/2} \\
& & \displaystyle \times \left[\frac{8R + 6M + 3\sqrt{8R^2 + 8MR + 6M^2}}{\big(4 + 3\sqrt{2}\big)\left(2R - 3M\right)}\right]^{1/2\sqrt{2}}.
\end{array}
\end{equation}
The conformal factor $\psi$ diverges at $r = 0$, which marks the black-hole ``puncture" in this trumpet geometry.  Expressions for components of the extrinsic curvature can be found in \cite{BauN07}.

As in Section \ref{sec:trans:iso}, we insert these expressions into Eqs.~(\ref{u_r}) and (\ref{u_t}) to find the radial and time components of the fluid four-velocity, and finally into Eq.~(\ref{v_r}) to find $v^r$.  The radial profiles of these variables are included in Fig.~\ref{Fig1} as dot-dashed lines.  We note that, unlike for the isotropic coordinates on a slice of constant Schwarzschild time $T$, all fluid variables remain finite on the horizon and extend smoothly into the black-hole interior, up to the limiting surface at $r = 0$.

\begin{figure}
\centering
\includegraphics[width=\textwidth]{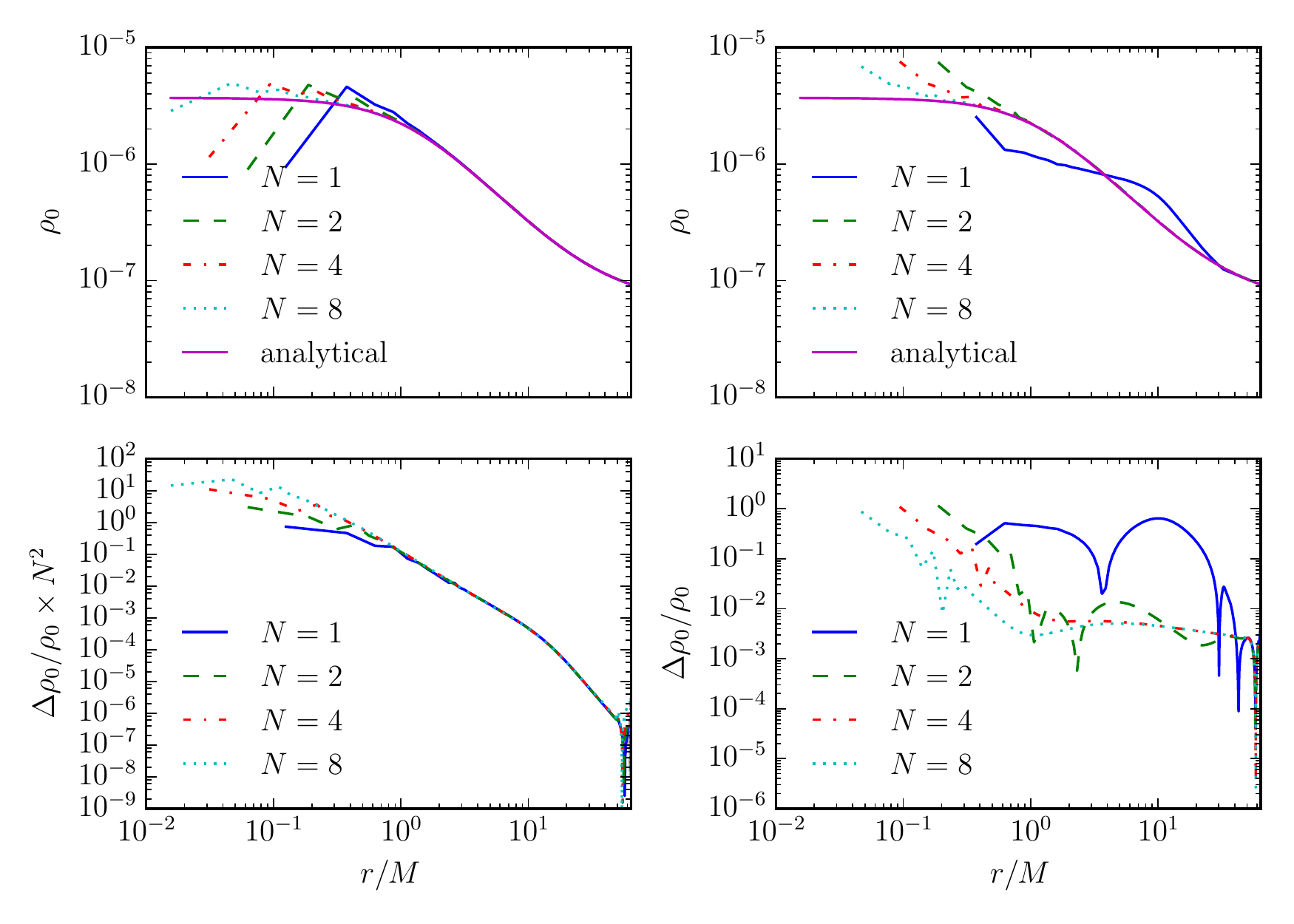}
\caption{The rest-mass density $\rho_0$ at $t = 63M$ for maximal trumpet initial data evolved with the 1+log slicing condition (\ref{1+log}) for $N_r = N \times 256$ radial grid points, where $N = 1$, 2, 4 and 8, and with the outer boundary imposed at $r_{\mathrm{max}} = 64M$.  Panels on the left show evolutions in the Cowling approximation, while those on the right show results for a fully dynamical evolution.  The top two panels show $\rho_0$ itself, while the bottom two show relative numerical errors $\Delta\rho_0/\rho_0$.  For the Cowling approximation (bottom left), these errors are multiplied with $N^2$ to demonstrate second-order convergence.}
\label{Fig3}
\end{figure}

We now evolve these data numerically using the slicing condition (\ref{1+log}), along with a (non-advective) Gamma-driver condition for the shift.  We evolve to a time $t = 63M$ and use $N_r = N \times 256$ uniform grid points, where $N = 1$, 2, 4 and 8, and impose the outer boundary at $r_{\mathrm{max}} = 64M$.

As a first test we evolve only the fluid, keeping the spacetime variables constant, see the left panels in Fig.~\ref{Fig3}.  In analogy to stellar perturbation calculations in which the perturbations of the gravitational potential are ignored, this approximation is sometimes referred to as the Cowling approximation, and we will use this term in the following.  The top left panel shows that, at $t = 63M$, the rest-mass density $\rho_0$ agrees well with the analytical solution except at a fixed number of about 6 or 8 grid points in the vicinity of the black-hole puncture at $r = 0$, where $\psi$ diverges.  As the resolution is increased, these grid points correspond to an increasingly small physical region inside the black hole.  In the bottom left panel of Fig.~\ref{Fig3} we show the relative error in the rest-mass density $\Delta\rho_0/\rho_0$.  For each resolution we scale $\Delta\rho_0/\rho_0$ by $N^2$; the resulting curves lie on top of one another (again except for a fixed number of grid points in the vicinity of the black-hole puncture), demonstrating second-order convergence for the fluid evolution.

We then relax the Cowling approximation and evolve the fluid variables self-consistently with the gravitational fields.  The top right panel in Fig.~\ref{Fig3} shows that, for sufficiently high resolution, the densities again agree well with the analytical solution, but we also see from the bottom right panel that the relative errors $\Delta\rho_0/\rho_0$ appear to level out at a few times $10^{-3}$, and do not further decrease with increasing resolution.  This behavior is not surprising, since the Bondi solution is only an approximate solution to Einstein's equations, i.e., it neglects the self-gravity of the fluid and the increase of the black-hole mass with accretion.  This approximation does not affect evolutions in the Cowling approximation; for full evolutions, however, it will lead to deviations of the evolved data from the initial data.  Given an accretion rate of $\dot{M}$, the black-hole mass $M$ should have increased by $\dot{M}t$ after a time $t$; ignoring this increase in the Bondi solution will lead to a relative error of about $\dot{M}t/M$.  For our value of $\dot{M} = 10^{-4}$, we therefore expect relative deviations of around $6 \times 10^{-3}$ at a time $t = 63M$, which is completely consistent with our numerical results.  For sufficiently high grid resolutions, when numerical errors have dropped below this level, the overall deviation of the numerical solution from the analytical solution is therefore dominated by this ``analytical'' deviation.

\begin{figure}
\centering
\includegraphics[width=0.75\textwidth]{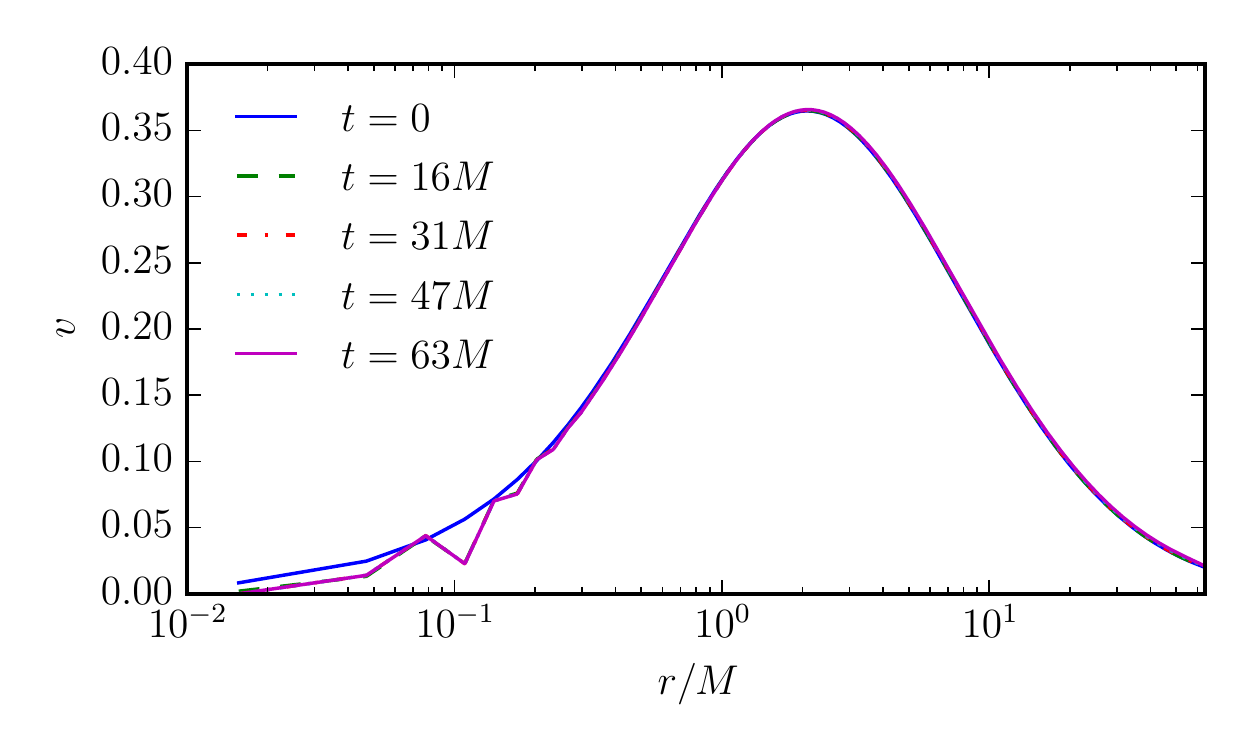}
\caption{Same as Fig.~\ref{Fig2}, but for maximal trumpet initial data.  Except for discrepancies at small radii inside the black hole, which are due to differencing across the puncture, profiles at different instants of time can barely be distinguished, demonstrating that even gauge-dependent quantities will remain time-independent in this setup of Bondi accretion.}
\label{Fig4}
\end{figure}

In Fig.~\ref{Fig4} we show radial profiles of the fluid velocity $v = -v^r$ at different instants of time, for a fully dynamical time evolution.  Except for slight discrepancies at small radii $r < M/2$ due to finite differencing across the black-hole puncture at $r = 0$, the profiles remain almost exactly constant, demonstrating that even gauge-dependent quantities remain time-independent for this choice of initial data and gauge conditions.  This behavior is very different from that of the analogous profiles for initial data on slices of constant Schwarzschild time shown in Fig.~\ref{Fig2}.

\subsection{Analytical trumpet coordinates}
\label{sec:trans:ana}

Finally, we briefly discuss results for a completely analytical family of trumpet slices of the Schwarzschild spacetime (see \cite{DenB14}, as well as \cite{DenBM14} for a generalization to Kerr black holes).  We choose the parameter $R_0$ defined in \cite{DenB14} to be $R_0 = M$.  The isotropic radius $r$ is then given by
\begin{equation} \label{r_of_R_ana}
r = R - M.
\end{equation}
In these coordinates the event horizon is located at $r = M$, and the limiting surface at $r = 0$ corresponds to an areal radius $R = M$.  We obtain remarkably simple expressions for the lapse,
\begin{equation} \label{lapse_ana}
\alpha = \frac{r}{r + M},
\end{equation}
the shift,
\begin{equation} \label{shift_ana}
\beta = \frac{Mr}{\left(r + M\right)^2},
\end{equation}
and the conformal factor,
\begin{equation} \label{psi_ana}
\psi = \left(1 + \frac{M}{r}\right)^{1/2}.
\end{equation}
Since we now have $\partial r/\partial R = 1$, we find from (\ref{u_r}) that $u^r = u^R$.  As in Sections \ref{sec:trans:max} and \ref{sec:trans:ana}, we compute $u^t$ from Eq.~(\ref{u_t}) and $v^r$ from Eq.~(\ref{v_r}).  Profiles of the fluid variables in these coordinates are included in Fig.~\ref{Fig1} as dotted lines.

\begin{figure}
\centering
\includegraphics[width=0.75\textwidth]{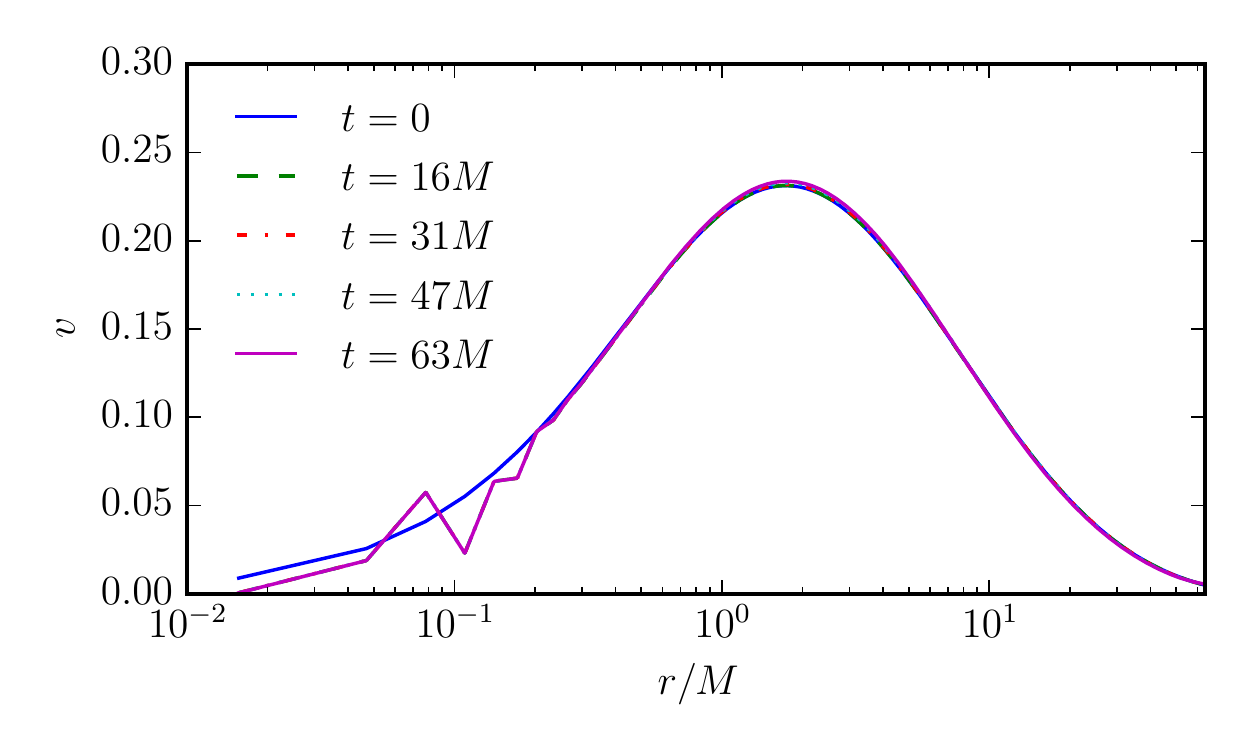}
\caption{Same as Figs.~\ref{Fig2} and \ref{Fig4}, but for analytical trumpet initial data evolved with the slicing condition (\ref{Ken_log}).  While numerical errors for these analytical trumpets are larger than those for the maximal trumpets of Fig.~\ref{Fig4}, profiles of $v$ still remain nearly unchanged, at least for $r$ sufficiently far from the black-hole puncture at $r = 0$.}
\label{Fig5}
\end{figure}

Data in these coordinates remain time-independent when evolved with a variation of the 1+log slicing condition (\ref{1+log}),
\begin{equation} \label{Ken_log}
\partial_t\alpha = -\alpha\left(1 - \alpha\right)K.
\end{equation}
As discussed in \cite{DenB14}, this slicing condition can lead to coordinate pathologies in general, but for spherically symmetric data we have nonetheless been able to carry out simulations using this condition.  While we have found that numerical errors are larger for these analytical trumpet slices than for the maximal trumpet slices of Section \ref{sec:trans:max}, profiles of $v$ still remain approximately constant over the course of the evolution, as shown in Fig.~\ref{Fig5}.  These results demonstrate that even this very simple setup can be used to test numerical simulations of fluid flow in self-consistently evolved black-hole spacetimes.

\section{Summary and discussion}
\label{sec:summary}

The Bondi solution, which describes spherically symmetric, radial fluid accretion onto a non-rotating black hole, provides a powerful test for relativistic hydrodynamics codes.  However, the Bondi solution is typically formulated in Schwarzschild coordinates, which, while convenient from an analytical point of view, cannot be implemented numerically.  Different coordinate transformations have therefore been used to cast the Bondi solution in coordinate systems that are more suitable for numerical evolution, but none of them extend smoothly into the black-hole interior and allow for a time-independent evolution with the moving-puncture coordinate conditions that have been so successful in numerical black-hole evolution calculations.

In this paper we transform the Bondi solution into two different trumpet coordinate systems, namely maximal trumpet and analytical trumpet coordinates.  In both of these coordinate systems, the Bondi solution extends smoothly into the black-hole interior and remains time-independent when evolved with moving-puncture coordinates.  Expressed in this way, the Bondi solution provides a powerful test for relativistic gravitohydrodynamics codes, allowing for direct comparisons with an analytical solution for fluid flow onto a black hole.  We demonstrate these features by performing convergence tests both inside and outside the black hole, while distinguishing between finite-differencing errors and errors that result from neglecting the self-gravity of the accreting fluid.

In this paper we focus on pure fluid accretion.  As demonstrated in \cite{DeVilH03}, however, the expressions for fluid flow remain unchanged in the presence of a (hypothetical) purely radial magnetic field.  Accordingly, this solution may also serve as a test for relativistic magnetohydrodynamics codes (see also \cite{GamMT03,DueLSS05b,MoeMFHNBLORS14}).

\ack

We would like to thank Ken Dennison for many helpful discussions and comments.  
This work was supported in part by NSF grant PHYS-1402780 to Bowdoin College.

\section*{References}
\providecommand{\newblock}{}

\end{document}